\newif\ifArxiv
\Arxivtrue

\ifArxiv
\documentclass[sigconf,10pt,nonacm]{acmart}
\else
\documentclass[sigconf,10pt,anonymous,review,nonacm]{acmart}
\fi

\newcommand{\memname}{Managed-Retention Memory}
\newcommand{\memacro}{MRM}

\ifArxiv
\title{\memname: A New Class of Memory for the AI Era}
\else
\title{Storage Class Memory is Dead, All Hail \memname : Rethinking Memory for the AI Era}
\fi

\author{\rm {Sergey~Legtchenko, Ioan~Stefanovici, Richard~Black, Antony~Rowstron, Junyi~Liu, Paolo~Costa, Burcu~Canakci, Dushyanth~Narayanan, Xingbo~Wu}}
\affiliation{%
  \institution{Microsoft Research}%
  \city{}
  \country{}%
}

\newcommand{\change}[1]{{\textcolor{black}{#1}}}
\newcommand{\para}[1]{{\bf \noindent #1\relax}}

\begin{abstract}
AI clusters today are one of the major uses of High Bandwidth Memory (HBM). However, HBM is suboptimal for AI workloads for several reasons. Analysis shows HBM is overprovisioned on {\em write} performance, but underprovisioned on density and read bandwidth, and also has significant energy per bit overheads. It is also expensive, with lower yield than DRAM due to manufacturing complexity. We propose a new memory class: \memname~(\memacro), which is more optimized to store key data structures for AI inference workloads. We believe that \memacro~may finally provide a path to viability for technologies that were originally proposed to support Storage Class Memory (SCM). These technologies traditionally offered long-term persistence (10+ years) but provided poor IO performance and/or endurance. \memacro~makes different trade-offs, and by understanding the workload IO patterns, \memacro~foregoes long-term data retention and write performance for better potential performance on the metrics important for these workloads.
\end{abstract}

\begin{document}

\maketitle

\section{Introduction}
To date the world has been very binary when it comes to storage: there are non-volatile and volatile storage technologies. DRAM in different forms (GDDR, HBM, LPDDR) is the dominant {\em volatile} memory storage technology. Data it stores is lost as soon as the energy source is removed. NAND block-oriented and NOR byte-addressable Flash are the most widely used examples of {\em non-volatile} memory storage. They do not need to be constantly powered to persist data. At the memory cell level, data volatility is expressed as {\em retention time}, which is the time that data is reliably stored without requiring a refresh. Flash cells have a retention time of 10+ years, but this comes at the cost of lower read and write throughput per memory cell than DRAM. These properties mean that DRAM is used as memory for processors, and Flash is used for secondary storage.

Several other memory technologies, like RRAM, MRAM~\cite{Meena2014OverviewOE, 10.5555/2536832} and PCM~\cite{10.1145/1555815.1555758}, all have the potential to offer non-volatility. They fall into a class of memory that is often referred to as Storage Class Memory (SCM) for servers. The recently discontinued Intel Optane\,/\,3D XPoint~\cite{optane} is an iconic representative of SCM, which aimed to overcome the IO limitations of Flash while being non-volatile. The dream was to replace DRAM by offering comparable IO performance and byte addressability, while also featuring 10+ year retention. However, all attempts to date have failed to displace DRAM due to the trade-offs. They failed to offer IO performance that is comparable to DRAM at lower (or same) costs as Flash due to the challenges of density and complex manufacturing processes. For main memory, persistence of data is not as important as IO performance. For general compute workloads, nobody wants to trade primary memory IO-performance for 10+ year data retention. These technologies also struggle with {\em endurance}, which refers to the number of write cycles a memory cell can support before it permanently degrades~\cite{10.1145/1555815.1555758}. Hence, SCM ended up being valuable for some use cases (e.g., embedded compute~\cite{weebit-market, intrinsic-semi}), but not for deployment in servers.

Ironically, we believe that the rise of Flash may have been something of a curse for memory innovation. Non-volatility is a key storage {\em device} property, but {\em at a memory cell} level it is quite misleading. For all technologies, memory cells  offer simply a retention time, which is a continuum from microseconds for DRAM to many years. The technologies that underpin SCM have been forced to be non-volatile, requiring their retention time to be a decade or more. Unfortunately, achieving these high retention times requires trading off other metrics such as write and read latency, energy efficiency and endurance~\cite{stt-mram,rram-endurance,7851470}.

Perhaps one reason why this has been viewed historically as binary, is that even with relaxed retention times, SCM technologies would not match DRAM on all metrics of importance {\em for general workloads}. However, {\em foundation models} (of which Large Language Models, or LLMs are a subset) have recently emerged as a new major workload with unique memory IO requirements~\cite{MSAIspend}. The tremendous scale and growth of foundation model training and inference require novel hardware approaches. Foundation model inference has different memory IO requirements to historical workloads. For example, a large fraction of the memory is used to store model weights, for which IO performance is critical for sequential {\em reads}, but much less important for {\em writes}. \change{Memory IO is sequential and predictable, and} given the energy challenges of AI clusters, energy per bit read is also an issue. The only technology today that can match the IO performance, energy and density is HBM. However, it is no panacea, and certain key stages in foundation model inference are memory not compute bound. Further, HBM is expensive and has significant yield challenges.

We think that there is an opportunity to rethink existing "non-volatile'' memory technologies for this new workload. We propose a new class of memory that we call {\em \memname}~(\memacro). \memacro~is different from volatile DRAM as it can retain data without power \change{and does not waste energy in frequent cell refreshes}, but unlike SCM, is not aimed at long-term retention times. As most of the inference data does not need to be persisted, retention can be relaxed to days or hours. In return, \memacro~has better endurance and aims to outperform DRAM (and HBM) on the key metrics such as read throughput, energy efficiency and capacity.

In the rest of this paper, Section~\ref{section:workload} first characterizes the foundation model workload characteristics and requirements. It then discusses the challenges and lack of optimality of HBM. Section~\ref{section:proposal} describes relevant emerging technologies. Finally Section~\ref{section:software} explores the broader systems implications of rethinking memory and introducing \memacro. We are explicitly not settling on a specific technology, instead highlighting an opportunity space. This is a call for action for those working on low-level memory cell technologies, through those thinking of memory controllers, to those designing the software systems that access the memory. Hail to a cross-layer collaboration for better memory in the AI era!

\section{Memory in the Foundation Model-era}
\label{section:workload}

The workload of a foundation model is quite different to traditional workloads. A foundation model is first {\em trained}, usually on a large cluster (e.g., 50,000+ AI accelerators), and the output is essentially a set of model weights. These weights are then deployed in production where they serve {\em inference} queries. Thousands or even millions of instances of the foundation models will be used but the scale of hardware per inference is much smaller (e.g., 4+ AI accelerators). It has been observed that both training and inference workloads are memory intensive\change{~\cite{yang2024protrain,agrawal2023sarathiefficientllminference}}. Training scale depends on the model size and is a one-off effort (often taking months), while the inference workload is demand-driven and served for a significant time period until the model weights are retired.

Training and inference have distinct memory access patterns and requirements, and are typically deployed on different clusters. As demand increases, we are expecting the inference infrastructure to dominate, and are hence focusing on the inference workload. More specifically, we consider foundation models that perform autoregressive token generation. An inference query is a sequence of input tokens, in response to which the foundation model generates a sequence of output tokens. A {\em context} is composed of all the tokens from the user and the corresponding responses generated by the model during the interaction. Having contexts as large as possible is desirable, as it improves the model's reasonning ability via its use of the self-attention mechanism~\cite{TransformerPaper}. However, in deployment, contexts have limited size and range from low 1000s to a few 10,000s tokens (depending on the model), and are primarily limited by the amount of memory available. Each inference query is computationally expensive and requires distributed computation across multiple AI accelerators.

Inference relies on three main in-memory data structures: {\em model weights}, the {\em KV cache}, and {\em model activations}. Of these, model weights and the KV cache use up the majority of the memory capacity~\cite{vLLM}.

The model weights (a matrix) have been key to expanding the capabilities of frontier foundation models; there has been an exponential growth in the size of the model weights with each generation of foundation model. Currently, large models have (well) over 500 billion weights, representing between 250~GB and over 1~TB of data depending on the weight quantization used. The weights are effectively a non-mutable data structure. The reference model weights are persisted in storage, while a replica is distributed across the AI accelerators in every inference cluster. There are a large number of foundation models today, but in practice a small number of the most popular ones are used at scale. All inference queries made to a given foundation model version (e.g., GPT4) use a copy of the same weights.

The KV cache supports the model's self-attention mechanism. It is a sequence of self-attention vectors that encode the model's understanding of the relationship between all the tokens in a context. Every time a new token is generated in a context, a vector is appended to the end of the corresponding KV cache. Each vector is typically a few MBs, so the KV cache usually grows to a few tens of GBs until the context size limit is reached.

Lastly, model activations are the transient tensors that are created and passed between the different layers during a forward pass of the network. They are typically an order of magnitude smaller than both the weights and the KV cache, and are only stored during the forward pass computation.

The KV cache is created during the {\em prefill} phase, when the first set of tokens is received from a user. Subsequently, in the {\em decode} phase the model iteratively generates response tokens. For that, at each iteration the KV cache is read entirely and sequentially, a new token is generated, and the corresponding self-attention vector is appended to the KV cache. KV caches leverage memory to reduce computation and are soft state: they are generated by the model, and can be re-computed if needed. However, the token rate per second is usually quite low (thus expensive) so caching and using the KV cache is usually preferable to recalculation.

During inference, the entire self-attention data and weights are read for every generated token, creating substantial bandwidth demand between memory and compute. At any given time, many inference requests are multiplexed over the same cluster, but all of them are for the same model. Each AI accelerator's memory thus contains a subset of the model weights, as well as several KV caches and activations that correspond to the working set of contexts. When a new model is deployed, the cluster stops accepting new requests, services ongoing ones, then loads weights for the new model.

To summarize, foundation model inference is mostly composed of very large, predictable memory reads, while writes are smaller and mostly append only. Exact memory ranges to be read are known in advance, and large fractions of the memory are not overwritten for long periods of time. Yet, despite being read-dominated, inference still requires write rates that are very high compared to storage workloads.

\subsection{The Curse of HBM}
\label{section:curse}

Today the majority of data used in an AI accelerator is stored on HBM, because all the data structures need to be repeatedly read at high bandwidth. Current AI accelerators can support very high main memory bandwidths, e.g., 8~TB/s for a single B200 GPU~\cite{b200}.  In addition, since weights and KV caches are large, AI accelerators require substantial HBM capacity. Hard engineering challenges need to be surmounted to achieve this, especially around energy usage. Signal loss over copper interconnect tracking at the required data rates means that the memory must be physically located (very) close to the compute die, typically co-packaged on the same interposer. The very wide interfaces, and high signal rates translate into more energy, and approximately a third of the energy usage for an AI accelerator is the memory.  HBM is used as it enables 3D-stacking of DRAM on the same \change{package} to boost on-package memory capacity, throughput, and minimize the distance of the memory cells from the AI accelerator. Current HBM products have 8-12 layers, for an aggregate 192~GB on a B200 package~\cite{b200}. Hence, HBM is used as it offers the highest throughput at the highest density with reasonable energy usage. However, even using HBM, a substantial part of every inference query is memory bound~\cite{DBLP:conf/isca/PatelCZSGMB24}.

Unfortunately, there is currently no viable alternative to HBM. Non-stacked DRAM does not have the required density, while NAND and NOR Flash memory are not fast enough and have low lifetime endurance especially at higher densities where multiple bits are stored per memory cell. Both lack the energy efficiency required in package.

It should be noted that HBM comes with several fundamental challenges. First, memory vendors are struggling to continue to scale the density. The per-layer scaling is struggling with challenges inherited from DRAM~\cite{memorywall}. So, the next generation of HBM (HBM4) is only expected to increase capacity per layer by 30\% compared to current HBM3e. Secondly, the 3D-stacking of DRAM both significantly reduces the yield of the manufacturing process and also leads to heat dissipation challenges, especially when tightly packaged with an AI accelerator die. Currently, the industry does not expect it to scale beyond 16 layers in the foreseeable future~\cite{micronroadmap}. 3D-stacking is extremely complex. These factors, combined with high demand, fueled by exponential growth of cloud infrastructure for foundation models, means that HBM accounts for a substantial fraction of an AI cluster's cost. This is unlikely to change in the foreseeable future, and AI clusters will remain dependent on HBM.

\subsection{A New Hope?}
\label{section:newhope}

Foundation model inference is very different from the general-purpose main memory workload for which DRAM was designed. First, it is extremely {\bf read-intensive}. For example, {\em each token} generated during decode requires reading {\em all} the weights, and the entire KV cache~\cite{patel2024splitwise}, for one self-attention vector write. Self-attention vector size is usually at most a few MBs~\cite{gpt3kv,llamakv}, while weights and KV caches are typically 10s of GBs, which imply read:write ratios of over 1000:1.

There are efforts to reduce the amount of data read during inference. For example, batching allows weight reuse across requests~\cite{agrawal2023sarathiefficientllminference}. However, batching is limited by latency requirements~\cite{agrawal2023sarathiefficientllminference}. Reuse of the KV cache across requests~\cite{prefixcaching} and KV cache compression~\cite{kvcacheCompression} are also used, but each has its limitations and even together they do not fundamentally change the heavily read-dominated nature of the workload.

Second, memory accesses are {\bf sequential} and {\bf predictable}. There are no in-place updates for weights or KV caches, and the same weights and KV cache are read iteratively for every foundation model response. Memory virtualization mechanisms have been proposed to address memory fragmentation~\cite{vLLM}, but even in that case, pages are read in the same order. Each page is typically over 10 vectors (typically several MBs to 10s of MBs) and is read sequentially~\cite{vLLM}. Furthermore, the mapping between virtual pages and physical addresses is typically static.

These properties suggest that most of the HBM capacity is used for data that has little use for the general-purpose properties HBM inherits from DRAM (random access, byte-addressability, comparable read and write performance). HBM is, in a sense, overprovisioned for the requirements of this foundation model inference workload. This overprovisioning leads to suboptimal cost and energy overheads.

It also raises the tantalizing question: if we correctly provision the memory to the workload, can we address this suboptimal cost and energy challenges for memory in inference clusters?



\section{The Memory Opportunity}
\label{section:proposal}

We posit that the combination of (i) the importance and scale of foundation model infrastructure, (ii) the large difference between the workload patterns of conventional server CPUs and that of AI accelerators, and (iii) the poor match of HBM to the workload, opens a field of computer architecture research in better memory for this application.

We now motivate that this opportunity is best addressed by a new type of memory, as opposed to DRAM, HBM or Flash. 
Flash cannot be used because it does not have enough endurance, even with Single Level Cells (SLC)~\cite{chang2007endurance}, and cannot satisfy the high throughput and energy efficiency requirements~\cite{papirla2009energy, 4505784}. The non-volatility of Flash is also unnecessary: the data is either persisted elsewhere (weights) or is soft state (KV caches, activations). 

On the other hand, some workload properties are close to ones typically exhibited by storage workloads. For example, byte addressability is not required, because IO is large and sequential. Similar to storage infrastructure, storage capacity and total cost of ownership (TCO)/TB are key metrics, on which DRAM and HBM are underperforming.

{\em Can \memacro~match AI cluster requirements?}  PCM, RRAM, and STT-MRAM have {\em read} performance and energy on par or better than DRAM, and potential for higher density and/or lower TCO/TB~\cite{ipek2010dynamically}. PCM was shipped at scale in Intel Optane devices, while RRAM and STT-MRAM have matured over the past few years, and are used for automotive, wearable and IoT applications~\cite{weebit-market,crossbar-reram,intrinsic-semi}. 

\begin{figure}
\centering
\includegraphics[width=3in]
{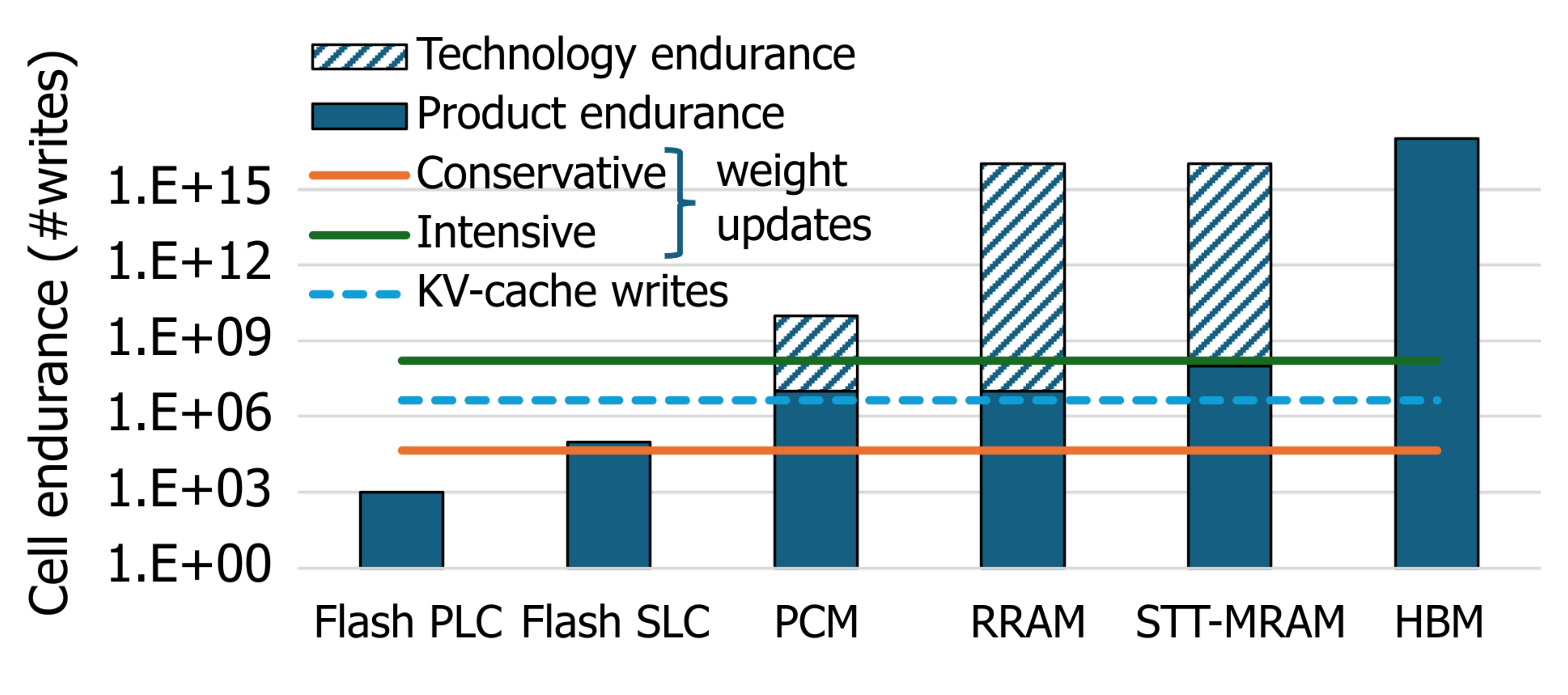}
\caption{\change{Endurance requirements for KV cache and model weights vs. endurance of memory technologies.}}
\vspace{-0.5cm}
\label{fig:endurance}
\end{figure}

These technologies have lower endurance than DRAM, and we now estimate the approximate endurance requirements for weight and KV cache writes. Weight updates are infrequent, bulk overwrites when the model is replaced. The update frequency is currently typically low (hours+), but could evolve as models diversify. We estimate the endurance required over 5 years for a conservative {\em hourly} update and an intensive {\em once per second} update. KV cache writes occur both during prefill and decode, one self-attention vector per context token. Prefill is typically higher throughput than decode, and we use the throughputs and median context lengths reported for the Llama2-70B model in Splitwise~\cite{patel2024splitwise}. For an expected lifetime of five years, we compute the number of KV cache writes, and infer the average number of writes per cell. 

Figure~\ref{fig:endurance} shows a comparison between endurance of existing memory/storage technologies and the workload endurance requirements. When applicable, we differentiate endurance observed in existing devices from the potential demonstrated by the technology. We use potential endurance from~\cite{Meena2014OverviewOE, 10.5555/2536832}, while device endurance is taken from device specifications and benchmarks (Intel Optane PCM~\cite{ optane-endurance}, Weebit RRAM~\cite{Molas2022HighTS} and Everspin STT-MRAM~\cite{7998174}). We observe that 1) HBM is vastly overprovisioned on endurance, and 2) existing SCM devices do not meet the endurance requirements but the underlying technologies have the potential to do so. We believe this is partly due to current devices being designed for non-volatility, which is achieved by trading off other important metrics such as write latency, energy efficiency or endurance~\cite{stt-mram,rram-endurance}. We see this as an opportunity to rethink existing memory technologies, currently used for SCM, specifically for AI workloads, by trading off non-volatility for other key metrics.


\section{Software Stack Implications}
\label{section:software}

In this section, we motivate why \memacro~is of interest to the computer systems community. Foundation models are becoming pervasive which leads to a diversification of the requirements: some use cases have tight latency SLAs (e.g., user-in-the-loop conversation), some are throughput hungry and heavily use batching, others are background best-effort jobs (e.g., meeting recap). The workload is becoming more complex, with vastly different input:output token ratios, expert models tailored for specific use cases, and dependencies on advanced augmentation mechanisms (e.g. RAG~\cite{zhao2024retrievalaugmentedgenerationaigeneratedcontent}). In addition to that, the resource-heavy nature of the workload and the cost of the hardware require hollistic and efficient orchestration. This is addressed by leveraging key OS mechanisms (e.g., virtual memory~\cite{vLLM}, power-aware scheduling~\cite{stojkovic2025tapasthermalpowerawarescheduling} or speculative execution~\cite{miao2023specinfer}), effectively building up towards a rack-scale OS for foundation model inference. In that context, the emergence of \memacro~brings a set of exciting challenges and opportunities to explore.

\para{Retention-aware data placement and scheduling.} \memacro~is unlikely to be a one-size-fits-all solution, and will co-exist with other types of memory, such as HBM for write-heavy data structures (e.g., activations), and LPDDR as a slower tier. Fine-grained understanding of lifetime and access patterns of the data will be required to lay out the data. The scheduler will need to track the data expiration times, and decide whether to refresh it or move it to another tier based on the state of the requests that depend on that data.

\para{Lightweight memory controllers.} There is potential to make the \memacro~controller extremely simple and energy efficient. The lack of random access requirements opens up a unique prospect of a block-level access {\em memory} controller, with implications on the software stack. Much of the functionality that is typically handled on the device, such as refresh and wear-levelling can be left up to a software control plane higher up in the stack, which is best-placed to make these decisions while satisfying global application requirements. This approach is akin to zoned storage interfaces for Flash~\cite{ZNS}.

\para{Dynamically Configurable Memory (DCM).} Since the control plane has cluster-level visibility over both applications and user workloads, it is also best-placed to {\em dynamically} decide the retention period needed for each data when it is written, effectively right provisioning the \memacro~to the workload. This is a fully-flexible instantiation of \memacro. At the hardware level, the memory controller would support writing at different durations and energies, allowing retention time to be programmed at runtime. The foundation model OS could then orchestrate optimal data refresh, wear-leveling, and garbage collection {\em at the cluster level}.

\para{Retention-aware error correction.} \memacro's relaxed retention requirements also raise an interesting question: how do we think about data integrity? Much of the data stored in \memacro~will either be durably stored elsewhere (e.g., weights), or be soft state (e.g., KV cache). As such, the requirements for persistence are not as stringent as for traditional storage systems. Nonetheless, the system still needs to enforce integrity in order to guarantee correctness of computation involving the data, and avoid frequent re-computation of soft state. Leveraging existing state-of-the-art error correction techniques for memory\cite{HBM3ECC} is a good start, however a large block-based \memacro~interface means that there is scope for considering error correction techniques that operate on larger code words and have less overhead~\cite{CodePerf}. Designing efficient error correction for \memacro~that meets the strigent latency and throughput requirements will be a fruitful area for open research.

\section{Related work}

The trade-offs between retention, endurance and write energy efficiency have been well studied both for STT-MRAM~\cite{5749716,6241517,7851483} and RRAM~\cite{10.1145/3656019.3676890, rram-endurance,5488761, Lammie_2021}. Leveraging this mechanism has been proposed to improve the energy efficiency of hybrid on-die CPU caches~\cite{5749716,6241517,7851483, 10.1145/3656019.3676890}. In contrast to our work, this strand of work focuses on general-purpose multi-core CPUs, and is hence addressing a different optimization problem. AI clusters have rack-scale energy and cooling requirements, and have a more complex set of memory tiers and interconnects, but more predictable workloads.

Stanford has recently started a 5-year project to address the anticipated upcoming increase in tiering and heterogeneity of main memory~\cite{stanford2024}. We share the same observation that the memory wall~\cite{memorywall} is a major challenge for key workloads, and is likely to lead to more memory heterogeneity due to lack of one-size-fits-all technology. While novel in data centers, this trend is common place in other applications. For example, the embedded world historically used ROM (Read Only Memory)~\cite{montangero1974approach,duhalde1995high} which was a write once read many technology, EPROM (Erasable Programmable Read Only Memory)~\cite{masuoka1987new} write few read many which was used to store programs and could be erased using UV light, and of course RAM. ROM and EPROM offered non-volatile storage, and careful design choices had to be made to best leverage the upsides of the different technologies.

There is ongoing effort to overcome the memory wall by tighly integrating memory and compute. This is done by either adding more memory onto the compute die~\cite{10609636,10631344}, or with in-memory computing (IMC)~\cite{verma2019memory}. Similar to our work, IMC is often aimed at AI workloads with either analog~\cite{analog} or digital~\cite{pim, 10477465} computation, and can be MRAM~\cite{9419046} or RRAM~\cite{9419046} based. Our work is orthogonal, because it aims to optimize the mainstream memory/compute model, instead of exploring a new paradigm.

Finally, there is substantial work on leveraging the heterogeneous memory access patterns in AI clusters. For example, it has been proposed to use CPU main memory for offloading idle KV caches~\cite{tangexploring,strati2024d}. The latest Nvidia's GB200 superchip has an integrated LPDDR5 controller for a higher capacity, slower memory tier~\cite{b200-specs}. This suggests that memory heterogeneity is going to be common place in AI clusters. Our work proposes to leverage more aspects of data access heterogeneity to maximize tokens generated per dollar. 

\section{Conclusion}

The emergence of AI workloads and their dependence on HBM memory has highlighted the limitations of HBM. AI inference workloads demand high read throughput, density, and energy efficiency, which HBM struggles to provide cost-effectively. We propose a new class of memory that can co-exist with HBM, \memname~(\memacro), which enables the use of memory technologies originally proposed for SCM, but trades retention and other metrics like write throughput for improved performance metrics crucial for these AI workloads. By relaxing retention time requirements, \memacro~can potentially enable existing proposed SCM technologies to offer better read throughput, energy efficiency, and density. We hope this paper really opens new thinking about innovation in memory cell technologies and memory chip design, tailored specifically to the needs of AI inference clusters.

%

\bibliographystyle{ACM-Reference-Format} 
\bibliography{paper}

\end{document}